\documentclass[onecolumn,showpacs,preprintnumbers,amsmath,amssymb]{revtex4}

\usepackage{epsf}
\usepackage{graphicx}  
\usepackage{dcolumn}   
\usepackage{bm}        

\newcommand{\be}{\begin{equation}}
\newcommand{\en}{\end{equation}}

\begin{document}

\preprint{arxiv:}

\title{Higgs Inflation on Braneworld}

\author{Hongsheng Zhang\footnote{Electronic address: hongsheng@shnu.edu.cn} }
\affiliation{ Center for
Astrophysics , Shanghai Normal University, 100 Guilin Road,
Shanghai 200234, China and State Key Laboratory of Theoretical Physics, Institute of Theoretical Physics, Chinese Academy of Sciences£¬Beijing, 100190, China
}
\author{Yi Zhang\footnote{Electronic address: zhangyia@cqupt.edu.cn} }
\affiliation{ College of Mathematics and Physics, Chongqing University o
f Posts and Telecommunications, Chongqing 400065, China and
Institute of Theoretical Physics, Chinese Academy of Science,
Beijing 100190, China
}

\author{Xin-Zhou Li \footnote{Electronic address: kychz@shnu.edu.cn} }
 \affiliation{ Center for
Astrophysics , Shanghai Normal University, 100 Guilin Road,
Shanghai 200234,China}

\date{ \today}

\begin{abstract}
We discuss a Higgs inflation model in the warped DGP braneworld background.
It generates reasonable primordial perturbations. At the same time with enhanced non-minimal coupling it overcomes
the severe problem in the Higgs inflation in 4 dimension, which says that the effective field theory
become invalid at an energy scale far below the energy scale for inflation exit. Furthermore we present the constraints
for the parameters confront to PLANCK and related observations. PLANCK low-l data almost fixes the inflation energy scale in this Higgs
inflation model with specific brane parameters.

\end{abstract}


\maketitle

\section{Introduction}

Recently the existence of Brout-Englert-Higgs-Guralnik-Hagen-Kibble boson (later, for convenience we adopt its traditional name Higgs boson) with a mass $m=125\sim 126$ Gev is confirmed at more than 5$\sigma$ confidence level \cite{higgs1}\cite{higgs2}. As the unique scalar particle in the standard model which endorses mass to
every massive particle, the Higgs boson achieves pivotal position in the standard model. Generally the inflation in the early universe is driven by a scalar field. A natural idea is that the Higgs boson drives the cosmic inflation. However we will find that it does not work, since the self-coupling coefficient for Higgs $\lambda=0.13$ due to LHC, which is much much higher than the perturbation required value $10^{-13}-10^{-14}$.

  Inflation can solve some problems in the standard big bang model, such as horizon, flatness, and surplus solitons. More significantly, it predicts a nearly
  scale-invariant perturbation spectrum, which is confirmed by the observations of cosmic microwave background. However, there are still
  several serious problems in the inflationary scenario. The most fundamental one is the physical nature of inflaton. In some sense inflation is
  only a paradigm. We still need to fill physics in this exponentially expanding frame. Usually, we assume a scalar beyond standard model to drive the inflation.
  Generally the properties of the scalar need to invoke new physics which we never see in territorial labs. If Higgs can drive the inflation, it is a competitive candidate since its physical foundation is now sound enough. As we have mentioned, a simple Higgs-driving-inflation is not successful. Thus a natural extension is to consider  a non-minimally coupling model. Recently, a non-minimally coupling
  Higgs inflation model is suggested \cite{non-higgs}. The quantum correction to this model is investigated in \cite{quantumhiggs1}\cite{quantumhiggs2}, for a review see \cite{reviewhiggs}.  The Lagrangian of this model reads,
\be
L_{nh}=\frac12 \mu^2f(\phi)R -\frac12\partial_{\mu} \phi\partial^{\mu}\phi-\frac{\lambda}{4}(\phi^2-v^2)^2.
\label{lnh}
\en
 Here $\phi$ is the Higgs field, $\mu$ and $R$ denote Planck mass and Ricci scalar respectively, $\lambda$ stands for the self-coupling constant, and $v=246$Gev is the vacuum expectation of Higgs field.
$f$ is a function of $\phi$,
\be
f(\phi)=1+\xi \frac{\phi^2}{ \mu^2},
\en
 where $\xi$ is a coupling constant. In a weak coupling limit $\xi\sim 1$ or $\xi<1$, Einstein-Hilbert term always dominates the non-minimal coupling term
 from the Planck era. Under this situation the model effectively comes back to general relativity with a small correction, and thus Higgs filed cannot drive the inflation. In the strong coupling
 limit for example $\xi>10^{34}$ Higgs field can drive a successful inflation. But in this case it almost decouples from all of the other fields and has a huge mass, which clearly
 contradicts with experiments. Some mediate value of $\xi\sim 10^4$ may generate a successful inflation model and, at the same time, does not contradict with the particle phenomenologies.
 Soon after this Higgs inflation model was proposed, several authors recognized a serious problem in this model. In this scenario, the inflationary phase exits at
 \be
 \frac{\phi}{\mu}\sim \frac{1}{\sqrt{\xi}},
 \label{phixi}
 \en
 while the effective field theory only makes sense at an energy scale no higher than
 \be
 \frac{\phi}{\mu}\sim \frac{1}{\xi},
 \en
 which is much below (\ref{phixi}) \cite{newphysics1}\cite{newphysics2}\cite{newphysics3}\cite{newphysics4}\cite{newphysics5}\cite{newphysics6}. This means that the action (\ref{lnh}) becomes invalid in the inflation phase. New physics must
 be aroused at that scale. This argument plagues the non-minimal coupling Higgs inflation in frame of general relativity. This result can be obtained both in Jordan frame and in Einstein frame.  We call it ``exorbitant exit energy problem".   Higgs is the unique scalar field
 in standard model and has an explicit potential form. It was recently found at LHC. To save the Higgs-driving-inflation model is an interesting topic. We shall show
 that it works well in frame of brane world gravity.

 In brane world scenario our universe is a 3+1 dimensional brane embedded in a
 higher dimensional spacetime (called bulk). Inspired by some early works such as Horova-Witten model \cite{hz},
  the standard model particles are assumed to be confined to a
3-brane, while gravity  propagates in both bulk and brane.
 In this picture, the success of standard model of particle physics is saved, but the physics related to gravity gets modified. Thus the achievements
 of the standard model are inherited and especially, the  properties of Higgs are the same as that of standard model. Among various brane universe models,
 the DGP (Dvali-Gabadadze-Porrati) model \cite{dgpmodel1}\cite{dgpmodel2}\cite{dgpcosmology1}\cite{dgpcosmology2}\cite{dgpcosmology3}\cite{dgpcosmology4}\cite{dgpcosmology5}\cite{dgpcosmology6}\cite{dgpcosmology7}\cite{dgpcosmology8} is one of the leading models in the studies of late time universe. If we introduce brane tension and bulk cosmological
 constant, we get a warped DGP model.  A warped DGP model has several interesting properties in the early universe \cite{wdgp1}\cite{wdgp2}. For instance, a scalar field with exponential potential
 can exit the inflationary phase spontaneously. Particle phenomenology is a significant topic in any serious brane world model.  The cosmology permitting non-minimal coupling between a scalar field confined to the brane and the induced Ricci scalar of the brane in brane world is studied in \cite{many1} \cite{many2}  \cite{many3}  \cite{many4}, and the evolution of a bulk scalar which non-minimally coupled to bulk Ricci scalar is studied in \cite{liuyx}. We shall show in this article that a warped DGP model can significantly change the energy scale of inflation, and thus overcomes
 the exorbitant exit energy problem in Higgs inflation picture.

 This article is organized as follows. In the next section we will present the theoretical frame the higgs inflation model on a warped DGP brane. In section III we analyse the permitted parameter space. In section IV we explore some numerical
 implications to compare with the observations, especially PLANCK satellite. In the last section we conclude this paper.

\section{Higgs dynamics in brane world}

The action of the generalized DGP model is written as
,
 \be
 S=S_{\rm bulk}+S_{\rm brane},
 \label{taction}
 \en
where
 \be S_{\rm bulk} =\int_{\cal M} d^5X \sqrt{-g_{(5)}}
\left[ {1 \over 2 \kappa_5^2} R_{(5)} + L_{\rm m}
\right], \label{bulk action}
 \en
and
 \be S_{\rm brane}=\int_{M} d^4 x\sqrt{-g} \left[
{f\over\kappa_5^2} K^\pm + L_{\rm brane}
\right].
 \label{brane_action}
 \en
Here $\kappa_5^2$ denotes the 5 dimensional Newton constant, $R_{(5)}$
and $L_{\rm m}$ are the 5 dimensional scalar curvature and the
matter Lagrangian in the bulk, respectively. $x^\mu
~(\mu=0,1,2,3)$ are the induced 4 dimensional  coordinates on the  brane,
$K^\pm$ is the trace of extrinsic curvature on either side of the
brane and $L_{\rm brane}$ is the effective
4 dimensional Lagrangian, which can be a generic functional of the
brane metric and matter fields.

 Now we consider a brane Lagrangian,
\be
L_{\rm brane}=  L_{nh} -  T,
\label{brane_action2}
\en
where $T$ denotes brane tension, and $L_{nh}$ is given by (\ref{lnh}).
 We assume that the bulk space includes only a
cosmological constant $\Lambda_{(5)}$. It can be treated as a generalized
version of the DGP model, which is obtained by
setting  $T=0$ as well as $\Lambda_{(5)}=0$, or a generalized version of RS model, which is obtained
by setting $\mu=0$ with non-vanishing $T$ and $\Lambda_{(5)}$.  We name this model warped DGP model \cite{wdgp1} \cite{wdgp2}. In this framework two types of actions, one is
the bulk action and the other is brane action, are included. A similar structure is investigated in the so called  Two
Measures Field Theory \cite{new1}\cite{new2}, where gravity and particle physics are intertwined in a
highly non trivial way.

 As in RS model, we define
 \be
\Lambda={1\over 2} (\Lambda_{(5)}+{1\over
6}\kappa_5^4T^2).
 \en
  Making variation with respect to the brane metric $g_{\mu\nu}$, we obtain the field equation. Assuming an FRW metric on the brane, we get the Friedmann equation,

\be
 H^2+{k\over a^2}={1\over 3\mu^2f(\phi)}\Bigl[\,\rho+\rho_0\bigl(1 +
\epsilon{\cal A}(\rho, a)\bigr)\,\Bigr],
 \label{fried}
 \en
 where as usual, $k$ is the constant curvature of the maximal symmetric
 3-subspace of the FRW metric and $\epsilon$ takes either $+1$ or $-1$. ${\cal A}$ is
defined by

\be
 {\cal A}=\left[{\cal A}_0^2+{2\eta\over
\rho_0}\left(\rho-\mu^2 {{\cal E}_0\over a^4}
\right)\right]^{1\over 2},
 \en
where
 \be
 {\cal A}_0=\sqrt{1-2\eta{\mu^2\Lambda\over \rho_0}},\ \
\eta={6m_5^6\over \rho_0\mu^2},
 \en
\be \rho_0=T+6{m_5^6\over \mu^2},
 \en
 and $a$ denotes the scale factor on the FRW brane, ${\cal E}_0$ labels a constant which is related to the contraction of bulk Weyl tensor,
 $m_5=\kappa_5^{-2/3}$ is the 5 dimensional Planck mass. In this article, we constrain ourself to a positive tension brane. Hence $\eta\leq 1$. Two degenerated cases are scalar-tensor gravity and warped minimally coupling DGP.
 One can check that the Friedmann equation (\ref{fried}) reduces to the case 4 dimensional scalar-tensor gravity \cite{self1} for $m_5=0$, and reduces to
 warped DGP model for $\xi=0$, respectively. These properties are also clear from the action form (\ref{taction}).

 Equation (\ref{fried}) is written in Jordan frame of the brane, in which scalar field is non-minimally coupled to gravity.  The explanation of observation results depends on  the ``ansats" of the frames\cite{sasa1}\cite{sasa2}. The frame which is more familiar to our experience is Einstein frame, in which the inflation physics is thoroughly studied.  Thus, it may be useful to see the results of Higgs inflation in the Einstein frame. Also it is more convenient to work in Einstein frame by using the full-blown formulism. To enter Einstein frame, we make a Weyl rescaling of the brane metric,
 \be
 g^{E}_{\mu\nu}=fg_{\mu\nu},
 \en
 where $E$ labels Einstein frame. The corresponding transformations for other quantities are listed as follows,
 \be
 \left(\frac{d\sigma}{d \phi}\right)^2=1/f+\frac{3}{2}\frac{\mu^2 f'^2}{f^2},
 \en
 \be
 T^{E}_{\mu\nu}=T_{\mu\nu}/f,
 \label{transmo}
 \en
    where $\sigma$ is a new field defined in Einstein frame, a prime stands for derivative with respective to $\phi$, $T_{\mu\nu}$ stands for the energy-momentum confined to the brane, which yields
    \be
    V(\sigma)^E=V(\phi(\sigma))/f^2,
        \en
        and
       \be
    T^E=T/f^2.
        \en
    $V=\frac{\lambda}{4}(\phi^2-v^2)^2$ is the Higgs potential. In Einstein frame, the action of the brane reads,
     \be
     S^E_{\rm brane}=\int_{M} d^4 x\sqrt{-g^E} \left[\frac{\mu^2}{2}R^E-\frac12\partial_{\mu} \sigma\partial^{\mu}\sigma-V^E(\sigma)-T^E
\right].
 \label{raction}
 \en
 Then Friedmann equation (\ref{fried}) becomes
 \be
 (H^E)^2+{k\over (a^E)^2}={1\over 3\mu^2}\Bigl[\,\rho^{E}+\rho^{E}_0\bigl(1 +
\epsilon{\cal A}(\rho^E, a^E)\bigr)\,\Bigr].
 \label{HE}
 \en
 Here, according to the transformation of energy-momentum (\ref{transmo}),
 \be
 \rho^E=\rho/f^2,
 \en
 \be
 \rho_0^E=\rho_0/f^2.
 \en
 Since ${\cal A}$ is a ratio between different densities, it is independent on conformal frames. We omit the superscript $E$ in the text what follows for we always work in Einstein frame
 without specific notations. As usual, we omit the curvature term and the Weyl term $\mathcal{E}_0$, since they are diluted rapidly in the inflationary phase. Furthermore, we adopt slow-roll approximation for a potential dominated Higgs.  Thus Friedmann equation (\ref{HE}) is further simplified to
 \be
 H^2=\frac{1}{3\mu^2}\left[V+\rho_0+\epsilon\rho_0\left({\mathcal A}_0^2+\frac{2\eta V}{\rho_0}\right)^{1/2}\right].
 \label{fv}
 \en
   Now we define the slow-roll parameters,
 \be
 \alpha=-\frac{\dot{H}}{H^2},
 \en
 \be
 \beta=-\frac{\ddot{H}}{\dot{H}H}.
 \en
  Substituting $H$ into the above equations, we reach,
  \be
  \alpha=\alpha_s\left[{1+\epsilon\eta ({\cal A}^2_0+2\eta x)^{-1/2}}\right]{\left[1+\frac1x+\epsilon \frac{({\cal A}^2_0+2\eta x)^{1/2}}{x}\right]^{-2}},
  \label{alpha}
  \en
  \be
  \beta=\beta_s\left[{1-\epsilon\eta^2 ({\cal A}^2_0+2\eta x)^{-3/2}}\right]{\left[1+\frac1x+\epsilon \frac{({\cal A}^2_0+2\eta x)^{1/2}}{x}\right]^{-1}},
  \en
  where $\alpha_s$ and $\beta_s$ are the corresponding slow-roll parameters in standard model,
  \be
  \alpha_s=\frac{1}{2}\frac{1}{V^2}\left(\frac{dV}{d\sigma}\right)^2,
  \en
  \be
  \beta_s=\frac{1}{V}\frac{d^2V}{d\sigma^2},
  \en
  and $x$ is the ratio $\rho/\rho_0$.
  The Friedmann equation (\ref{fv}) looks exactly the same as the Friedmann equation in warped DGP model \cite{wdgp1}. However, we stress that all the quantities
  in this equation are written in Einstein frame. For simplicity, we define,
  \be
  B=1+\epsilon \eta({\cal A}^2_0+2\eta x)^{-1/2},
  \label{defineB}
  \en
  \be
  I=1+\frac1x+\epsilon \frac{({\cal A}^2_0+2\eta x)^{1/2}}{x}.
  \en
   The e-folds number $N$ reads,
  \be
  N=\int_{t_i}^{t_e} Hdt,
  \en
 where $t_i$ and $t_e$ denote the epoches when presently observed universe exits Hubble radius and inflation ends, respectively. Substituting $H$ in (\ref{HE}) in the above equation, we arrive at $N$ expressed by $\phi$ and $I$,
  \be
  N=\frac34 I \xi (\phi_i^2-\phi_e^2)/\mu^2.
  \en
   Here $\phi_i$ and $\phi_e$ are the values of $\phi$ at $t_i$ and $t_e$, respectively. Comparing this formula with the corresponding one in standard model, we see that $I$ is an e-folds amplifier.  $\alpha=1$ when $\phi=\phi_e$, thus
   we obtain
   \be
   \phi_e=\left(\frac43\right)^{1/4}B^{1/4}\mu(\xi I)^{-1/2},
   \label{termi}
   \en
   by using (\ref{alpha}). Then we get
   \be
   \phi_i^2=\mu^2\left(\frac{2}{\sqrt{3}}\frac{\sqrt{B}}{\xi I}+\frac{4N}{3I\xi}\right).
   \label{phii}
   \en
   The slow roll parameters read
   \be
   \alpha=\frac{4}{3}\frac{B\mu^4}{\xi^2I^2\phi^4},
   \label{alpha1}
   \en
   and
   \be
   \beta=-\frac{4}{3}\frac{\mu^2}{\xi I\phi^2}\left[{1-\epsilon\eta^2 ({\cal A}^2_0+2\eta x)^{-3/2}}\right]\left(1-\frac{\mu^2}{\xi\phi^2}\right),
   \label{beta1}
   \en

  The PLANCK normalization of primordial scalar perturbation at $k=0.05$ Mpc$^{-1}$ requires \cite{planck1}\cite{planck2}\cite{planck3},
  \be
  A_s=2.2\times 10^{-9}.
  \en
  Here $A_S$ can be calculated as follows. In Einstein frame, the perturbation formulism just follows the standard model. The primordial curvature perturbation ${\cal{R}}_k$ reads
  \be
   {\cal{R}}_k=-\frac{H}{\dot{\sigma}}\delta \sigma_k,
   \en
   where $k$ is a perturbation wave number. We note that in Jordan frame where the inflation field is $\phi$ the relation between ${\cal{R}}_k$ and $\phi_k$ will be different \cite{noh}. Then its spectrum can be expressed by $H$ and $\dot{\sigma}$ at the epoch of Hubble radius exit ($k=aH$),
   \be
   A_s=\frac{H^4}{4\pi^2\dot{\sigma}^2}.
  \en
   Substituting the slow roll approximation
   \be
   \frac{dV}{d\sigma}+3H\dot{\sigma}=0,
   \en
   into the above equation, we reach,
   \be
   A_s=\frac{9}{4\pi^2}H^6\left(\frac{dV}{d\sigma}\right)^{-2}.
   \en

  We note that the amplitude of scalar perturbation is normalized at $k=0.002$ Mpc$^{-1}$ in previous observations, including COBE, WMAP, and BOOMERanG etc. PLANCK
  normalizes this quantity at a smaller scale since its resolution is significantly increased at small scale. This difference affects our judgements of e-folds before inflation exit.
  For example if we use $N=60$ in previous models normalized by WMAP, we should set $N=56.7$ for this the same model normalized by PLANCK. However, we are not certain about the exact e-folds. The uncertainty is 10 or more. So we can still set $N=60$ in ordinary discussions.
  By using the equations (\ref{fv}) and (\ref{phii}), we reach
  \be
  A_s=\frac{\lambda I}{128\pi^2\xi^2}\left(2\sqrt{\frac{B}{3}}+\frac{4N}{3}\right)^2.
  \label{as}
  \en
  Now we take a look at the energy bound for the validity of the effective field theory in non-minimally coupling warped DGP model. Since the Higgs field only non-minimally couples to
  the induced 4 dimensional Ricci tensor, the local physics exactly follows the non-minimally coupling case in standard 4 dimensional case, which has been studied in several articles via several different methods \cite{newphysics1}\cite{newphysics2}\cite{newphysics3}\cite{newphysics4}\cite{newphysics5}\cite{newphysics6}. The result is that the effective theory holds only below the scale
  \be
  \phi_b\sim \mu/{\xi}.
  \label{phi_b}
  \en
  From (\ref{termi}), the inflation ceases at
  \be
  \phi_e\sim \mu/{\sqrt{\xi}},
  \en
  in standard model with $I=B=1$. For a non-minimally coupling Higgs inflation in standard model, the magnitude of scalar perturbation requires $\xi\sim 10^{4}$. It is clear that the success of Higgs inflation model is plagued by this argument, because of the exorbitant exit energy problem, as we have mentioned in the introduction section. We shall see that this problem is well resolved  in frame of warped DGP.

   \section{Parameter space analysis}
 The most conservative view requires the inflation must cease before nuclear synthesis, for we have enough evidences that nuclear synthesis occurs at a decelerating universe. Less conservative considerations lead that inflation should exit at an energy scale higher than Tev scale \cite{tev}. We adopt this ``less conservative" point.
 The potential in Einstein frame,
 \be
 V=\frac{\lambda}{4}\frac{\phi^4}{(1+\xi\phi^2/\mu^2)^2},
\en
  where we have omitted the term $v^2$, since it is far below the energy for inflation exit. We consider the energy region where
  \be
  \xi\phi^2/\mu^2>>1.
  \en
  The region of $\xi\phi^2/\mu^2<<1$ effectively comes back to the minimal coupling case, which is a not successful model of inflation. Then
  $V$ is further simplified to
  \be
  V=\frac{\lambda\mu^4}{4\xi^2}\left(1-\frac{2\mu^2}{\xi\phi^2}\right).
  \en
 In almost all the time of inflationary phase, $V$ is in fact a constant $V=\frac{\lambda\mu^4}{4\xi^2}$. We require
 \be
 V\geq (1{\rm Tev})^4,
 \en
 which yields
 \be
 \xi<10^{30},
 \en
 for $\lambda=0.13$. It is a huge number, but still lower than the strong coupling case by 4 orders \cite{strong1}\cite{strong2}\cite{strong3}\cite{strong4}. The energy scale for inflationary phase exit must be lower than
 the failure scale of effective field theory,
 \be
  \left(\frac43\right)^{1/4}B^{1/4}\mu(\xi I)^{-1/2}<\mu/{\xi},
    \en
 that is,
 \be
 \frac{B^{1/4}}{\sqrt{I}}<\xi^{-1/2}\left(\frac43\right)^{-1/4}
 \label{tuichu}
 \en
   In the low energy limit with $\frac{\xi\phi^2}{\mu^2}<<1$, the conformal factor $f$ goes to 1 and the theory is effectively equal to the minimal coupling case. The nonminimal coupling warped DGP and minimal coupling warped DGP share the same low energy behaviours, which can determine the region of the parameters in this model. We invoke some results of the minimal coupling warped DGP without demonstrations. For detailed demonstrations of these results, see \cite{wdgp1}. In the case DGP model with $\eta =1$ and ${\cal A}_0=1$, the low energy phenomenology of cosmology requires $\rho_0\sim (10^{-3}{\rm ev})^4$ for either branch $\epsilon=\pm 1$. For warped DGP model with with $\lambda \ne 0$ and $\Lambda_{(5)} \ne 0$ however,  $\rho_0\geq (1{\rm Mev})^4$ also can satisfy all observations.

   First, we consider the case of original DGP model, in which $\rho_0 \sim (10^{-3}{\rm ev})^4$. $V$ should be much higher than $\rho_0$ in inflationary phase, which says $x>>1$. Then we have $I=1$. The slow roll parameter becomes,
   \be
   \alpha=\alpha_s \frac{B}{I^2}=\alpha_s\left[1+\epsilon x^{-1/2}(1-\frac12 x^{-1})\right].
   \en
   $x$ is a very large number, and hence the correction to the standard model is negligible, either in the branch $\epsilon=1$ or $-1$. So the exit energy scale keeps the same as that of standard model, which is not helpful to the hinge of exorbitant exit energy problem.

   Second we consider a warped DGP model, i.e., we turn on the 5 dimensional cosmological constant and brane tension, for which $\rho_0> (1{\rm Mev})^4$ can satisfy all late time observations. Case I : $x>>1$, which implies a small $\rho_0$. Then $I=1$. (\ref{tuichu}) becomes,
   \be
    B^{1/4}<\xi^{-1/2}\left(\frac43\right)^{-1/4}.
    \label{Bu}
    \en
    From (\ref{as}), we have the upperbound of $B$ satisfies,
    \be
    \frac{4B_u}{3I^2}\left(2\sqrt{\frac{B_u}{3}}+\frac{4N}{3}\right)^2=\frac{128\pi^2 A_s}{\lambda}.
    \en
    We take the $A_s=2.2\times 10^{-9}$ from PLANCK, $\lambda=0.13$ from LHC, and e-folds $N=56$. The numerical
    result is $B_u=2.5\times 10^{-9}$. We see that $B$ must be a tiny number. This result  fulfills our physical intuition from (\ref{termi}), that is, $B$ should be very small to lower the exit value of $\phi$ if $I=1$. However from (\ref{defineB}), $B\sim 1$ under the condition $x>>1$ branch, which contradicts to such a tiny $B$. Then we consider case II with a large $\rho_0$,ie, $x<<1$. We assume $x<<1$ throughout the inflation phase we considered. In this case (\ref{tuichu}) becomes,
    \be
    I>\frac{2}{\sqrt{3}}(1+\epsilon\eta)^{1/2}\xi.
    \en
     From (\ref{as}), we have the lowerbound of $I_l$ satisfies,
    \be
    I_l=\frac{\lambda (1+\epsilon\eta)}{96\pi^2 A_s}\left(2\sqrt{\frac{1+\epsilon\eta}{3}}+\frac{4N}{3}\right)^2.
    \en
    The numerical result is $I_l=4\times 10^8$ if we set the parameter $\eta=0$. $\eta$ always takes a value in $\eta\in (0,1]$. So the lower bound of $I\sim 10^8$. The upperbound of inflation occurs at an energy scale
    \be
    x_u=(I-1)^{-2}\left(I-1+\epsilon\eta-\sqrt{(I-1+\epsilon\eta)^2-(I-1)^2(1-{\cal{A}}_0^2)}\right)\sim 10^{-8}.
    \label{xie}
    \en
    The lowerbound of $x$ reads,
    \be
    x_l\sim 10^{-15},
    \label{xie}
    \en
    according to our fundamental assumptions about the energy scale of inflation exit.   One can easily verify that the inflation ends at the scale equal or lower than the upperbound of effective field theory in this subcase. Thus we overcome the exorbitant exit energy problem in frame of warped DGP.

      Now we present a concrete example from the only reasonable subcase in case II of warped DGP. We assume $\eta$ is negligible and $I=4\times 10^8, B=1, \xi=3.4\times 10^8, N=56$. Under this parameter set,
      we get $\alpha=2.1\times 10^{-4}$, $\beta=-0.019$.
       The spectrum index of scalar perturbation reads,
       \be
       n_s=1-6\alpha+2\beta=0.962,
       \en
      which is well consistent with the latest PLANCK result.
     The ratio of amplitude of tensor to scalar reads,
     \be
     r=16\alpha=3.4\times 10^{-3},
     \en
     which is also in perfect accordance with PLANCK observations.

\section{Confronting to Planck}

We constrain the parameters in this model by using the recent Planck low-l data \cite{planck1}\cite{planck2}\cite{planck3}. From the discussion in the last section, we see that only case II ($x<<1$) in warped DGP can overcome the exorbitant exit energy problem in non-minimal Higgs inflation model. We shall show that the constraint results confirm this analytical demonstrations.

 In figures 1 and 2, we assume $x\ll 1$,  $\epsilon=1$, $\lambda=0.1312$, $N=56$, $\Psi=\sqrt{\xi}\frac{\phi}{\mu}$. In figure 1, we display the key quantities in this model confront to the key observational quantities in Planck observations, in which $x$ is the energy scale when the wave length of perturbation with wave number $k=0.05$Mpc$^{-1}$ equals Hubble radius in the inflationary phase. In fact in slow-roll inflation the energy density is almost a constant, hence $x$ denotes inflation energy scale. The integrated likelihood of 1$\sigma$ and $2\sigma$ levels for $x$ is $x=1.44^{+0.32+0.37}_{-0.29-0.29} \times 10^{-10}$. One sees that this is exactly below the upperbound which we present in last section,  $ x_u\sim 10^{-8}$. To confirm this point we show the constraint results with variable parameters in figure 2. In every case the inflationary energy scale $x$ is higher than the lowerbound and lower than the upperbound, which is well consistent with our analytical demonstrations. The integrated likelihoods of 1$\sigma$ and $2\sigma$ levels for the ratio of scalar to tensor perturbations  $r$  in figure 1 is $r=1.06_{-2.13-2.13}^{+21.94+23.82}\times 10^{-2}$, which get deep into the permitted region by Planck. From figure 2, one sees that the shape of contour plots for different parameters are almost exactly the same. The central value for the energy scale $x$ shifts to a smaller one when $A_0$ decreases or $\eta$ increases.

 \begin{figure}
 \centering
  {\includegraphics[width=6in]{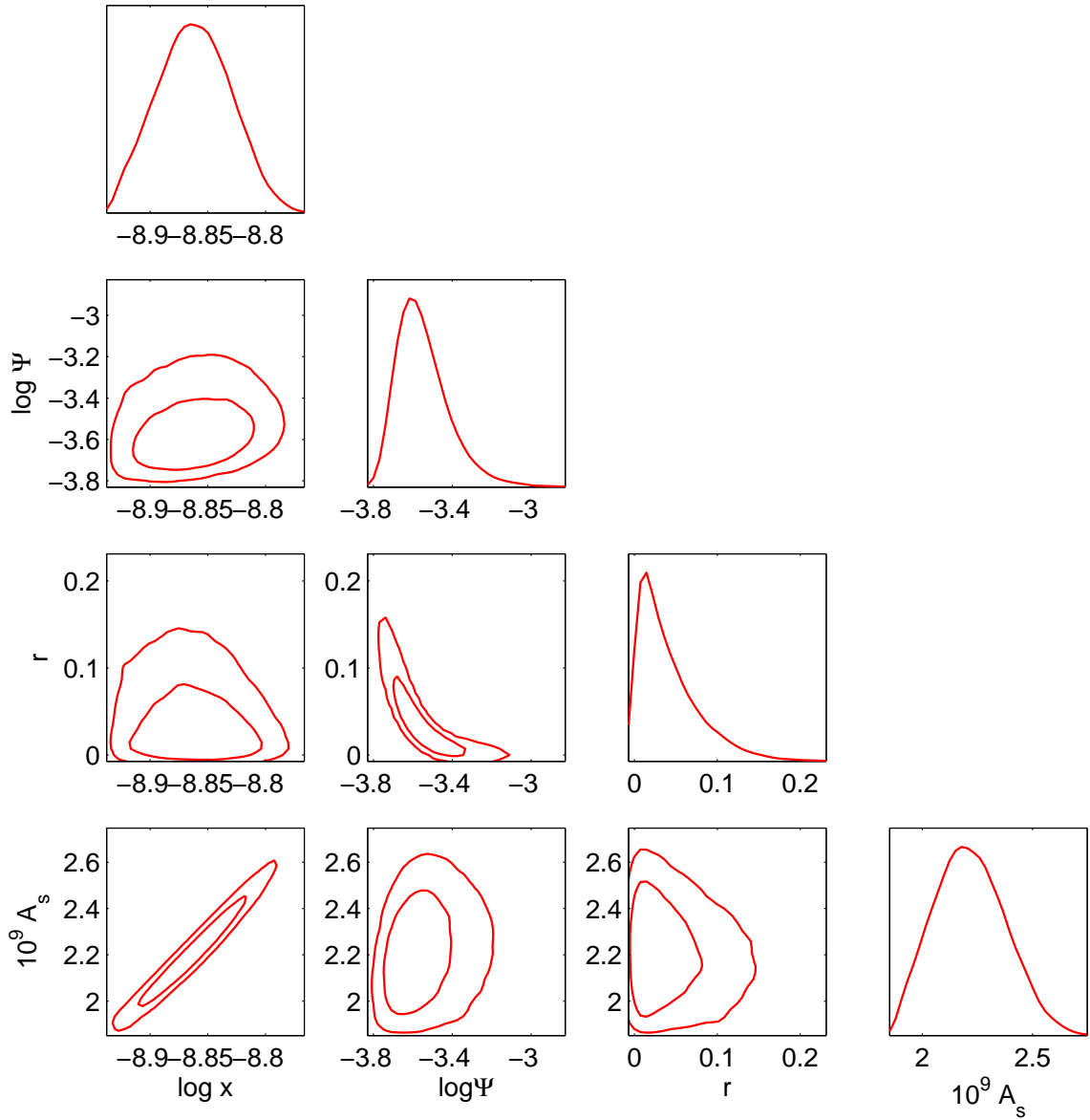}}
\caption{Multiple constraint results for $x, \Psi, r, A_s$. From the left to the right, from the top to bottom: likelihood of $x$, $1\sigma$ and $2\sigma$ contours on the plane
$\log \Psi$---$\log x$, likelihood of $\log\Psi$,  $1\sigma$ and $2\sigma$ contours on the plane $r$---$\log x$, $1\sigma$ and $2\sigma$ contours on the plane $r$---$\log \Psi$, likelihood of $r$, $1\sigma$ and $2\sigma$ contours on the plane $A_s$---$\log x$, $1\sigma$ and $2\sigma$ contours on the plane $A_s$---$\log \Psi$, $1\sigma$ and $2\sigma$ contours on the plane $A_s$---$r$,
 and likelihood of $A_s$.}
\label{tri}
\end{figure}

 \begin{figure} \centering
  {\includegraphics[width=3.5in]{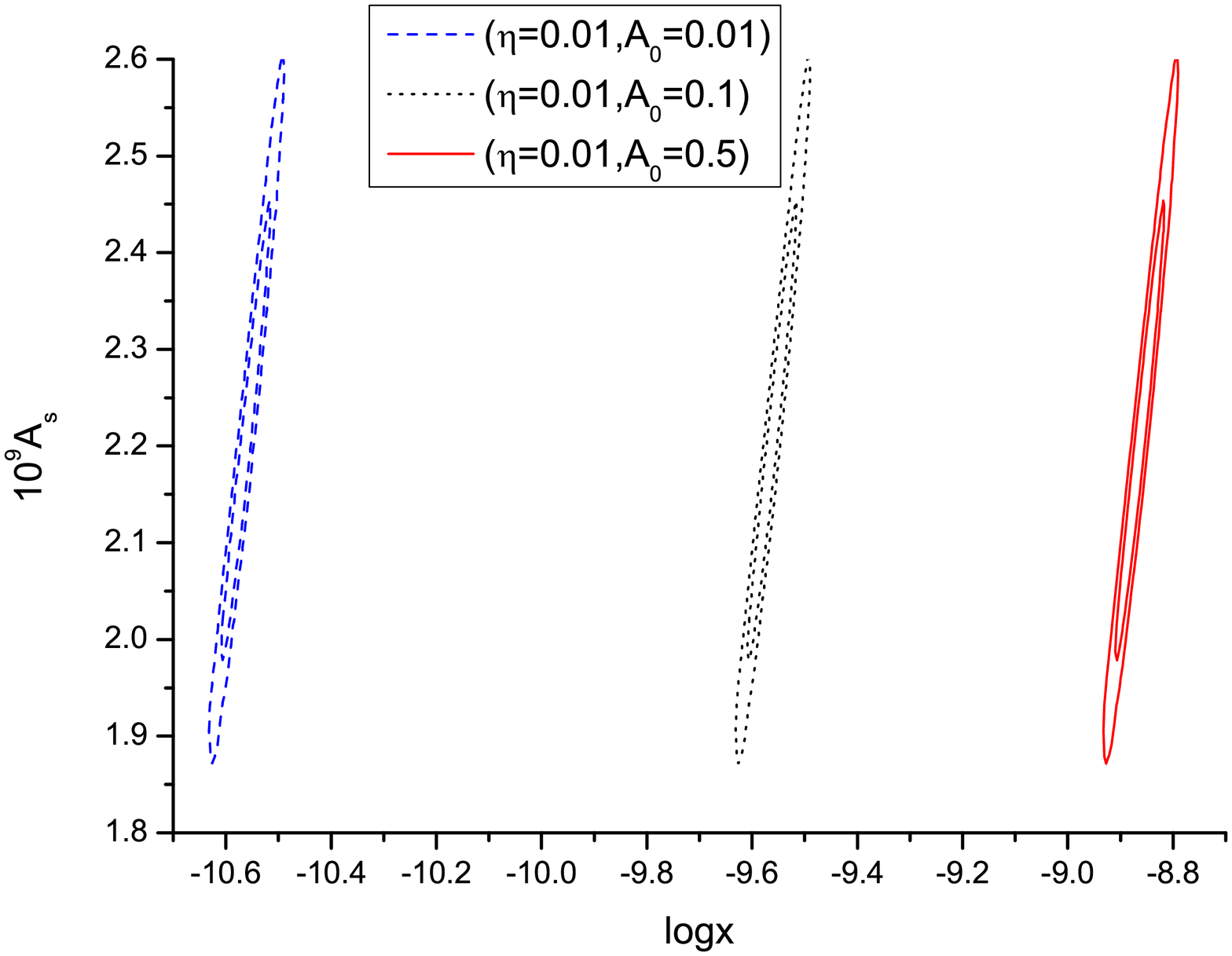}\label{eta1}}
   {\includegraphics[width=3.5in]{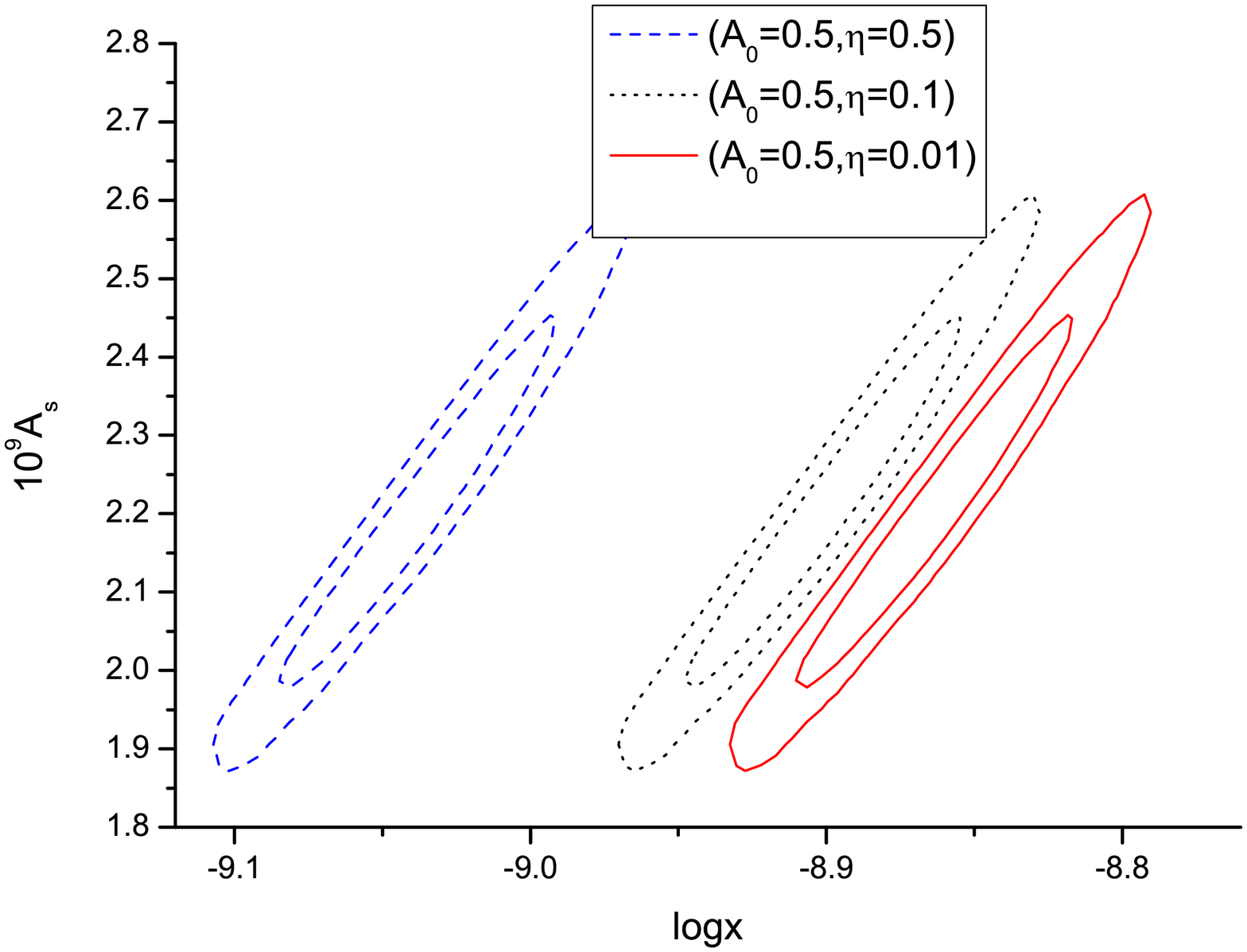}\label{A01}}
\caption{ Left panel: The $1\sigma$ and $2\sigma$ contours on the plane $A_s$---$\log x$ for varying $\eta$. Right panel: The $1\sigma$ and $2\sigma$ contours on the plane $A_s$---$\log x$ for varying $A_0$.}
\end{figure}


\section{Conclusion and discussion}

We find that Higgs inflation runs very well in warped DGP model. The exorbitant exit energy problem is overcome. Higgs is the unique scalar field in standard model of particle physics. If It will be the most economic model if Higgs itself can drive inflation. Unfortunately, exorbitant exit energy problem occurs in such a simple model. We consider the ``secondary economic model", ie, an inflation driven by Higgs on a brane. With help of the extra freedoms of the brane model, $\xi$ can be enhanced  to $10^8$, and thus the inflation exit at a reasonable energy scale. The reasonable energy scale dwells at $10^3$GeV to $10^9$GeV. The Planck low-l result presents rather tight constraints on the parameters. The inflation scale is almost be fixed with specific braneworld parameters. Reversely, if we can find the energy scale for inflation by other way, it is very helpful to determine the model parameters.

{\bf Acknowledgments:} This work is supported by the Program for Professor of Special Appointment (Eastern Scholar) at Shanghai Institutions of Higher Learning, National Education Foundation of China under grant No. 200931271104, National Natural Science Foundation of China under Grant Nos. 11075106,  11275128, 11175270, 11005164, 11073005 and 10935013.

\end{document}